\documentclass[12pt,a4paper]{iopart}

\usepackage{graphicx}
\usepackage{datetime}
\usepackage{color}
\usepackage{mathrsfs} 
\usepackage{dsfont}
\usepackage{xspace}
\usepackage{iopams}  

\usepackage[bookmarks=true,
            bookmarksopen=true,colorlinks=false]{hyperref}

\newcommand{\eqref}{\eref}
\newcommand{\dd}{\mathrm{d}}
\newcommand{\ee}{\mathrm{e}}
\newcommand{\ii}{\mathrm{i}}

\newcommand{\CC}{\mathds C}

\newcommand{\E}{\mathcal E}

\newcommand{\Al}{\mathcal A_1}
\newcommand{\Ar}{\mathcal A_2}
\newcommand{\Hp}{{\mathcal H_\mathrm{p}}}
\newcommand{\Hf}{{\mathcal H_\mathrm{f}}}
\newcommand{\p}{_\mathrm{p}}
\newcommand{\f}{_\mathrm{f}}
\newcommand{\Y}{{\mathbf Y}}
\newcommand{\J}{{\mathbf J}}
\newcommand{\B}{{\mathbf B}}
\newcommand{\M}{{\mathbf M}}
\renewcommand{\H}{{\mathbf H}}
\newcommand{\1}{{\mathbf 1}}
\newcommand{\0}{{\mathbf 0}}
\newcommand{\C}{{\mathbf C}}
\newcommand{\EE}{{\mathbf E}}

\renewcommand{\Im}{\mathrm{Im}}
\newcommand{\Sth}{\ensuremath{\mathbb S^3}}
\newcommand{\Sone}{\ensuremath{\mathbb S^1}}
\newcommand{\Stwo}{\ensuremath{\mathbb S^2}}

\begin{document}

\title[SGGTN solutions in Einstein-Maxwell theory]
{Smooth Gowdy-symmetric generalised Taub-NUT solutions in Einstein-Maxwell theory}

\author{J\"org Hennig}
\address{Department of Mathematics and Statistics,
           University of Otago,
           PO Box 56, Dunedin 9054, New Zealand}
\eads{\mailto{jhennig@maths.otago.ac.nz}}

\begin{abstract}
We introduce a new class of inhomogeneous cosmological models as solutions to the Einstein-Maxwell equations in electrovacuum. The new models can be considered to be nonlinear perturbations, through an electromagnetic field, of the previously studied  `smooth Gowdy-symmetric generalised Taub-NUT solutions' in vacuum. Utilising methods from soliton theory, we analyse the effects of the Maxwell field on global properties of the solutions. In particular, we show existence of regular Cauchy horizons, and we investigate special singular cases in which curvature singularities form.
\\[2ex]{}
\end{abstract}


\section{Introduction\label{sec:intro}}

Penrose's strong cosmic censorship hypothesis \cite{Penrose1969, MoncriefEardley1981, Chrusciel1991} (see also \cite{Rendall2005, Ringstrom2009, Isenberg2015}) is considered to be one of the most important unresolved issues in mathematical general relativity and cosmology. If this conjecture is true, then the set of spacetimes containing Cauchy horizons is, in some sense, a very small subset of all solutions to the Einstein equations. This would imply that spacetimes in which causality and determinism break down could practically not occur in nature, and hence be very desirable from a physical point of view.

As a first step towards a better understanding of cosmic censorship, it is useful and interesting to study the `undesired' solutions in detail and to investigate
properties of spacetimes \emph{with} Cauchy horizons. Some exact solutions, as well as abstract solution families, are already known. 

The typical prototype example is the Taub-NUT solution \cite{Taub1951, NUT1963, MisnerTaub1969}, which was characterised by Misner as \emph{`a counterexample to almost anything'}  \cite{Misner1967}. This model is rather special as it describes a spatially \emph{homogeneous} solution with a four-dimensional group of isometries. 

Another interesting example is given by the Kerr solution, since the region between the event and Cauchy horizons has a reinterpretation as an unpolarised Gowdy spacetime with $\Sone\times\Stwo$ topology \cite{Obregon}. Due to a compactification of the Boyer-Lindquist time coordinate, the solution in this interpretation does have closed timelike curves in the `black hole sector' (the region outside the event horizon). An analysis of the interior of much more general dynamical rotating vacuum black holes without any symmetry was recently presented in \cite{DafermosLuk}. These investigations are particularly relevant in the context of certain formulations of the strong censorship conjecture.

Existence of a large class of (generally) \emph{inhomogeneous} cosmological solutions with $U(1)$ isometry group, the \emph{generalised Taub-NUT spacetimes}, was shown by Moncrief \cite{Moncrief1984}. 
The starting point for this paper is 
a related class of solutions, which was introduced in \cite{BeyerHennig2012}: the \emph{smooth Gowdy-symmetric generalised Taub-NUT} (SGGTN) solutions. Like Moncrief's spacetimes, the SGGTN solutions are vacuum solutions with spatial three-sphere topology $\Sth$. However, they require less regularity (smoothness instead of analyticity), but they contain more symmetries (Gowdy symmetry with two spacelike Killing vectors). Existence of these models was shown in \cite{BeyerHennig2012, Hennig2016b}, and several families of exact solutions in this class were constructed with methods from soliton theory in \cite{BeyerHennig2014, Hennig2016b, Hennig2016a}. For a related application of soliton methods to Gowdy-symmetric cosmological models with spatial topology $\mathbb S^2\times\mathbb S^1$, we refer to \cite{AnsorgHennig2010}, and a general overview of strong cosmic censorship in the class of Gowdy spacetimes can be found in \cite{Ringstrom2010}.

The aim of this paper is to include a nontrivial additional field by generalising the class of SGGTN solutions from vacuum to electrovacuum. In particular, we extend the soliton methods used in \cite{BeyerHennig2012}, which allows us to study global properties of SGGTN solutions with a Maxwell field present. Firstly, however, we summarise some key features of the vacuum case.

Vacuum SGGTN models are solutions to the Einstein vacuum equation with spatial $\Sth$ topology, and they admit two commuting Killing vectors $\xi$ and $\eta$. Moreover, they have a past Cauchy horizon (by definition) and, generally, a second, future Cauchy horizon (as a consequence). These horizons are Killing horizons whose generators are linear combinations of the two Killing vectors. The original considerations in \cite{BeyerHennig2012} assumed that the past horizon is generated by $\xi$ alone, which corresponds to a horizon with closed orbits. In \cite{Hennig2016b} this was extended to generators of the form $\xi-a\p\eta$, $a\p=\rm{constant}$, which generally corresponds to non-closed orbits. It turned out that this generalisation is relatively straight-forward: the new situation can essentially be reduced to the case where $\xi$ generates the horizon by modifying one of the axes boundary conditions and allowing a different behaviour of one of the metric potentials at the horizon. Based on this observation, we will simplify the calculations in the present paper by assuming that the past horizons are always generated by $\xi$.

Note that, according to the interesting considerations in \cite{MoncriefIsenberg2018, PetersenRacz2018}, horizons with closed generators are in some sense `more typical', because they can occur for spacetimes with one symmetry only, while non-closed generators (under certain assumptions) imply existence of at least two symmetries. In case of the SGGTN solutions (which do have two symmetries), there generally is at least one horizon with non-closed generators: even if we assume that the past horizon is generated by $\xi$ alone, the future horizon is generated by a linear combination $\xi-a\f\eta$, where $a\f\neq 0$ in general.

Furthermore, the SGGTN vacuum solutions have two functional degrees of freedom. We can choose two free smooth functions at the past horizon (subject to a periodicity condition, in case that the past horizon is supposed to be generated by $\xi$ only), which uniquely determine the corresponding solution. In general, the solutions, which always start from a regular past horizon, also develop a regular future horizon. Only in special `singular cases' (corresponding to initial data satisfying an additional condition), a curvature singularity forms at the future boundary of the maximal globally hyperbolic development.

For more details about the vacuum case we refer to \cite{BeyerHennig2012, Hennig2016b}.

In the present paper, we study how SGGTN solutions behave if we include an electromagnetic field. For that purpose, we consider solutions to the electrovacuum Einstein-Maxwell equations.
In Sec.~\ref{sec:field} we use the field equations to derive necessary conditions for local existence of electrovacuum SGGTN solutions, and we determine the available degrees of freedom at the past horizon. Afterwards, in Sec.~\ref{sec:ErnstLP}, we first reformulate the field equations in terms of a pair of Ernst equations and, subsequently, in the form of an associated linear matrix problem. By integrating the linear problem at the symmetry axes and horizons,  we derive global properties of the solutions in Sec.~\ref{sec:integration}. Finally, in Sec.~\ref{sec:discussion}, we summarise and discuss our results.

\section{Field equations and local existence\label{sec:field}}

We start from the coordinates $(t,\theta,\rho_1,\rho_2)$ as used in \cite{BeyerHennig2012} with line element
\begin{equation}\label{eq:metric0}
  \dd s^2=\ee^M(-\dd t^2+\dd\theta^2)+R_0\left[\sin^2\!t\,\ee^u (\dd\rho_1+Q \dd\rho_2)^2+\sin^2\!\theta\,\ee^{-u} \dd\rho_2^2\right],
\end{equation}
where $u$, $Q$ and $M$ are functions of $t$ and $\theta$ alone, and $R_0$ is a positive constant. The angles $\rho_1$, $\rho_2$ take on values in the regions
\begin{equation}\label{eq:rhodomain}
 \frac{\rho_1+\rho_2}{2}\in(0,2\pi),\quad
 \frac{\rho_1-\rho_2}{2}\in(0,2\pi).
\end{equation}
In these coordinates, the two Killing vectors for Gowdy symmetry are simply given by  coordinate vectors, $\xi=\partial_{\rho_1}$ and $\eta=\partial_{\rho_2}$.
Initially, the solutions are defined in the `Gowdy square' with
\begin{equation}
 t\in(0,\pi),\quad \theta\in(0,\pi).
\end{equation}
Note that the boundaries $\theta=0,\pi$ of the Gowdy square are the symmetry axes $\Al$ and $\Ar$, and the boundaries $t=0,\pi$ correspond to the past and future Cauchy horizons $\Hp$ and $\Hf$, respectively. 

The construction of the coordinates and the requirement that hypersurfaces $t=\mathrm{constant}$ have 3-sphere topology imply the following axes boundary conditions\footnote[1]{As mentioned in the introduction, we assume throughout the article that $\Hp$ is generated by $\xi=\partial_{\rho_1}$. Otherwise, the axes boundary values of $\ee^{M+u}$ would be given by different constants \cite{Hennig2016b}.} \cite{BeyerHennig2012},
\begin{equation}\label{eq:axescond}
 \Al:\quad Q=1,\quad \ee^{M+u}=R_0,\qquad
 \Ar:\quad Q=-1,\quad \ee^{M+u}=R_0.
\end{equation}

In a next step, we replace $t$ and $\theta$ by
\begin{equation}
 x=\cos\theta,\quad y=\cos t.
\end{equation}
Then the line element becomes
\begin{equation}\fl\label{eq:metric}
   \dd s^2 =\ee^M\left(\frac{\dd x^2}{1-x^2}-\frac{\dd y^2}{1-y^2}\right)+R_0\left[(1-y^2)\,\ee^u 
   (\dd\rho_1+Q\,\dd\rho_2)^2+(1-x^2)\,\ee^{-u} \dd\rho_2^2\right].
\end{equation}
The choice of $x$ and $y$ instead of $t$ and $\theta$ is useful for extending solutions through the Cauchy horizons and for obtaining simpler versions of the Ernst equations and the linear problem below.

The energy-momentum tensor for the electromagnetic field is
\begin{equation}\label{eq:Tij}
 T_{ij}=\frac{1}{4\pi}(F_{ki}F^k{}_j-\frac14 g_{ij}F_{kl}F^{kl}),
\end{equation}
where, for our symmetries, the field tensor $F_{ij}$ is given in terms of an electromagnetic 4-potential of the form $(A_i)=(0,0,A_3,A_4)$ by
\begin{equation}
 F_{ij}=A_{i,j}-A_{j,i}. 
\end{equation}
The resulting Einstein-Maxwell equations with respect to our coordinates were derived in \cite{Hennig2016a}. For completeness, we list them again.

We have a second-order equation for $u$,
\begin{eqnarray}
     \fl\nonumber
     &&(1-x^2)u_{,xx}-(1-y^2)u_{,yy} -\frac{1-y^2}{1-x^2}\,\ee^{2u}
      \left[(1-x^2)Q_{,x}^{\ 2}-(1-y^2)Q_{,y}^{\ 2}\right]
       -2xu_{,x}+2yu_{,y}+2\\
     \fl&& \nonumber
      \quad = \frac{2}{R_0}\left(\frac{\ee^u}{1-x^2}
       \left[(1-x^2)(A_{4,x}-QA_{3,x})^2-(1-y^2)(A_{4,y}-QA_{3,y})^2\right]
       \right.\\
     \fl&&  \qquad\left.
       -\frac{\ee^{-u}}{1-y^2}\left[(1-x^2)A_{3,x}^{\ 2}-(1-y^2)A_{3,y}^{\ 2}\right]\right),\label{eq:ueq}
    \end{eqnarray}
a second-order equation for $Q$,
\begin{eqnarray}
    \nonumber
    \fl&&(1-x^2)Q_{,xx}-(1-y^2)Q_{,yy}+2(1-x^2)Q_{,x}u_{,x}
    -2(1-y^2)Q_{,y}u_{,y}+4yQ_{,y}\\
    \fl&&\quad  =-\frac{4\ee^{-u}}{R_0(1-y^2)}
    \left[(1-x^2)(A_{4,x}-QA_{3,x})A_{3,x}-(1-y^2)(A_{4,y}-QA_{3,y})A_{3,y}\right],\label{eq:Qeq}
   \end{eqnarray}
and two first-order equations for $M$,
\begin{eqnarray}\label{eq:Mx}
   \fl M_{,x} &=& -\frac{1-y^2}{2(x^2-y^2)}\Bigg[
    x(1-x^2)u_{,x}^{\ 2}+x(1-y^2)u_{,y}^{\ 2}-2y(1-x^2)u_{,x}u_{,y}
     \nonumber\\
   \fl &&
   +2\frac{x^2+y^2-2x^2y^2}{1-y^2}u_{,x}-4xyu_{,y}-4x\nonumber\\
   \fl &&
   +\frac{1-y^2}{1-x^2}\ee^{2u}
   \Big(x(1-x^2)Q_{,x}^{\ 2}+ x(1-y^2)Q_{,y}^{\ 2}
   -2y(1-x^2)Q_{,x}Q_{,y}\Big)\nonumber\\  
   \fl &&
   +\frac{4\ee^u}{(1-x^2)R_0}\Big[\frac{1-x^2}{1-y^2}\ee^{-2u}
   \Big(x(1-x^2)A_{3,x}^{\ 2}+x(1-y^2)A_{3,y}^{\ 2}
    -2y(1-x^2)A_{3,x}A_{3,y}\Big)\nonumber\\
   \fl &&
   +x(1-x^2)(A_{4,x}-QA_{3,x})^2+x(1-y^2)(A_{4,y}-QA_{3,y})^2\nonumber\\
   \fl &&
   -2y(1-x^2)(A_{4,x}-QA_{3,x})(A_{4,y}-QA_{3,y})\Big]\Bigg],\label{eq:Meq1}
  \end{eqnarray}
 \begin{eqnarray}\label{eq:My}
  \fl  M_{,y} &=& \frac{1-x^2}{2(x^2-y^2)}\Bigg[
    y(1-x^2)u_{,x}^{\ 2}+y(1-y^2)u_{,y}^{\ 2}-2x(1-y^2)u_{,x}u_{,y}\nonumber\\
   \fl &&
   +4xy u_{,x}-2\frac{x^2+y^2-2x^2y^2}{1-x^2}u_{,y}-4y\nonumber\\
   \fl &&
   +\frac{1-y^2}{1-x^2}\ee^{2u}
   \Big(y(1-x^2)Q_{,x}^{\ 2}+ y(1-y^2)Q_{,y}^{\ 2}
   -2x(1-y^2)Q_{,x}Q_{,y}\Big)\nonumber\\  
   \fl &&
   +\frac{4\ee^u}{(1-x^2)R_0}\Big[\frac{1-x^2}{1-y^2}\ee^{-2u}
   \Big(y(1-x^2)A_{3,x}^{\ 2}+y(1-y^2)A_{3,y}^{\ 2}
    -2x(1-y^2)A_{3,x}A_{3,y}\Big)\nonumber\\
   \fl &&
   +y(1-x^2)(A_{4,x}-QA_{3,x})^2+y(1-y^2)(A_{4,y}-QA_{3,y})^2\nonumber\\
   \fl &&
   -2x(1-y^2)(A_{4,x}-QA_{3,x})(A_{4,y}-QA_{3,y})\Big]\Bigg].\label{eq:Meq2}
  \end{eqnarray}

Moreover, Maxwell's equations lead to the following two equations,
\begin{equation}\label{eq:Max1}
   \left[\ee^u(A_{4,x}-QA_{3,x})\right]_{,x}
   =\left[\frac{1-y^2}{1-x^2}\,\ee^u(A_{4,y}-QA_{3,y})\right]_{,y},
\end{equation}
\begin{equation}\label{eq:Max2}
  \fl \left[\frac{1-x^2}{1-y^2}\,\ee^{-u}A_{3,x}-Q\ee^u(A_{4,x}-QA_{3,x})\right]_{,x}
  \!\! =\left[\ee^{-u}A_{3,y}-\frac{1-y^2}{1-x^2}Q\,\ee^u(A_{4,y}-QA_{3,y})\right]_{,y}\!\!.
\end{equation}

From the above Einstein-Maxwell equations, we can derive boundary conditions at the horizons and axes, and determine which quantities can be specified at the past horizon and which are already fixed by the equations. Together with the axes boundary conditions, this leads to necessary conditions for local existence of regular solutions (in a neighbourhood of the past horizon).

Firstly, we derive conditions at the symmetry axes $x=\pm1$. If we multiply \eqref{eq:ueq} by $1-x^2$ and perform the limit $x\to\pm1$, then we  obtain that a sum of two squares must vanish. This leads to two conditions: the function $Q$ must be constant on the axes [which is automatically true, since $Q=\pm1$ at $\mathcal A_{1/2}$, see \eqref{eq:axescond}], and $A_4\mp A_3$ is constant at $\mathcal A_{1/2}$.
Moreover, from \eqref{eq:Meq2}, we obtain for $x\to\pm1$ that $M+u$ is constant on both axes. However, this is again automatically true, since $\ee^{M+u}=R_0$ at the axes [cf.\ \eqref{eq:axescond}].

Secondly, we derive conditions at $\Hp$, where we require that the potentials and their derivatives are bounded. If we multiply \eqref{eq:ueq} by $1-y^2$, then we obtain for $y\to 1$ that $A_3$ is constant at the past horizon. Without loss of generality, we could even choose $A_3=0$ there, since only derivatives of the 4-potential are relevant. Moreover, from \eqref{eq:Meq1}, we obtain in the limit $y\to1$ that $M-u=\mathrm{constant}$.

Next we investigate which data can be chosen at the past horizon $y=1$. For the quantities $u$, $Q$, $A_3$, $A_4$, which satisfy PDEs of second order in $y$, we would normally expect that function values and $y$-derivatives can be specified at a surface $y=\mathrm{constant}$. However, due to the degeneracy of the equations at $\Hp$ ($y=1$), this is not the case for most of our potentials, with the exception of $A_3$. 

From \eqref{eq:Max2} we obtain in the limit $y\to1$ an equation that contains the second derivative $A_{3,yy}$. Hence $A_{3,y}$ is \emph{not} fixed by this equation. Consequently, in addition to the constant (and physically irrelevant) value of $A_3$, we can also choose $A_{3,y}$ at $\Hp$. Then \eqref{eq:ueq} leads for $y\to 1$ to an equation that does not contain second-order $y$-derivatives. Instead, it can be solved for $u_{,y}$ so that these values are fixed. Hence we can only prescribe the function values of $u$, but not the $y$-derivative. Similarly, from \eqref{eq:Max1} we obtain the values of $A_{4,y}$, so that we can only choose $A_4$, but not the $y$-derivative. Finally, from \eqref{eq:Qeq} we obtain a formula for $Q_{,y}$. Therefore, we can again only choose function values but not the $y$-derivative.

Furthermore, the axis boundary conditions must be satisfied, in particular, at the points where $\Al$ and $\Ar$ approach the past horizon, i.e.\ at the points
\begin{equation}
 A:\quad t=0,\ \theta=0\ (x=y=1)
\end{equation}
and
\begin{equation}
 B:\quad t=0,\ \theta=\pi\ (x=-1,\ y=1).
\end{equation}
As mentioned above, there we must have that $\ee^{M+u}=R_0$. Together with the condition that $M-u$ is constant on $\Hp$, we obtain the following (where the subscripts $A$ and $B$ refer to function values at the points $A$ and $B$),
\begin{eqnarray}
   R_0\ee^{-2u_A}&=&\ee^{M_A+u_A}\ee^{-2u_A}=\ee^{M_A-u_A}=\ee^{M-u}\big|_{\Hp}\nonumber\\
   &=&\ee^{M_B-u_B}=\ee^{M_B+u_B}\ee^{-2u_B}=R_0\ee^{-2u_B}. 
\end{eqnarray}
It follows --- exactly as in the vacuum case --- that $u_A=u_B$ must hold. This condition ensures that the first-order $M$-equations \eqref{eq:Meq1}, \eqref{eq:Meq2} have a solution for which the boundary condition $\ee^{M+u}=R_0$ is satisfied on \emph{both} axes (even though we can only choose \emph{one} integration constant).

Combining the previous observations, we see that we can find initial data by choosing a constant value for $A_3$ and smooth functions for $u$, $Q$, $A_{3,y}$ and $A_4$ on $\Hp$ such that
\begin{equation}\label{eq:conditions}
 u_A=u_B,\quad Q_A=1,\quad Q_B=-1.
\end{equation}
Note that $M$ cannot be specified on $\Hp$, since it is completely fixed by the first-order equations \eqref{eq:Meq1}, \eqref{eq:Meq2}  together with the axes boundary conditions \eqref{eq:axescond}.

According to the above derivation, regularity of all functions in a vicinity of the past horizon necessarily implies that we are restricted to the above choices of initial data subject to \eqref{eq:conditions}. Hence these are necessary conditions for local existence of regular solutions (with smooth potentials). Interestingly, in the vacuum case, the similarly derived conditions turned out to be not only necessary but also sufficient. The same can be expected for the above conditions in the electrovacuum case, but a rigorous proof of this statement would require extension of the underlying Fuchsian methods from the vacuum case to the present situation, which is beyond the scope of this paper.

Note that, instead of prescribing the metric and electromagnetic potentials at $\Hp$ as described above, we can also give the initial function values of the corresponding Ernst potentials, which will be introduced in the next section. 

\section{Ernst formulation and the linear problem\label{sec:ErnstLP}}

We intend to derive global properties of electromagnetic SGGTN solutions. For that purpose, we assume that data at the past horizon have been chosen for which a solution exists in some (arbitrarily small) neighbourhood of $\Hp$. Most likely, this is true for all data of the form discussed in the previous section. But at the very least, we can be sure that the set of permissible initial data is not empty, because exact solutions are known to exist: a homogeneous solution was given by Brill \cite{Brill1964} (see also Sec.\ 12.4 in \cite{GriffithsPodolsky}), and an inhomogeneous model was constructed in \cite{Hennig2016a}.

Once local existence is ensured (due to an appropriate choice of data), global existence in the entire Gowdy square --- with possible exception of the future boundary $t=\pi$ ($y=-1$) --- follows immediately. This was shown by Chru{\'s}ciel for the vacuum case \cite{Chrusciel1990}, and the argument carries over to electrovacuum  \cite{ChruscielPrivate}. However, it is not immediately clear how the solution behaves at $t=\pi$, which could be the location of a curvature singularity, a future Cauchy horizon, or something else. In order to investigate this behaviour, we reformulate the field equations in terms of the Ernst equations and the associated linear problem. 

The Ernst formulation of the field equations is obtained by introducing two complex Ernst potentials $\Phi$ and $\E$ as follows.
Firstly, we define auxiliary functions $f$ and $a$ in terms of the Killing vectors,
\begin{eqnarray}
  \label{eq:f}
  f &=& \frac{1}{R_0}g(\partial_{\rho_2},\partial_{\rho_2})
        = Q^2\ee^u(1-y^2)+\ee^{-u}(1-x^2),\\
  \label{eq:a}
  a &=& \frac{g(\partial_{\rho_1},\partial_{\rho_2})}
             {g(\partial_{\rho_2},\partial_{\rho_2})}
        = \frac{Q}{f}\ee^u(1-y^2),
\end{eqnarray}
and a function $\beta$ via
\begin{eqnarray}
 \label{eq:betax}
 \beta_{,x} &=& \frac{f}{1-x^2}(A_{3,y}-aA_{4,y}),\\
 \label{eq:betay}
 \beta_{,y} &=& \frac{f}{1-y^2}(A_{3,x}-aA_{4,x}).  
 \end{eqnarray}
Secondly, we construct the first Ernst potential,
\begin{equation}\label{eq:Phi}
 \Phi = \frac{1}{\sqrt{R_0}}(A_4+\ii\beta).
\end{equation}
Thirdly, we define another function $b$ via
\begin{eqnarray}
  \label{eq:ax}
  a_{,x} &=& \frac{1-y^2}{f^2}\left[b_{,y}
             +\ii(\bar\Phi\Phi_{,y}-\Phi\bar\Phi_{,y})\right],\\
  \label{eq:ay}
  a_{,y} &=& \frac{1-x^2}{f^2}\left[b_{,x}
             +\ii(\bar\Phi\Phi_{,x}-\Phi\bar\Phi_{,x})\right].
\end{eqnarray}
Finally, we define the second Ernst potential, 
\begin{equation}\label{eq:E}
  \E   = f+|\Phi|^2+\ii b.
\end{equation}

Now we can reformulate the Einstein-Maxwell equations in terms of $\Phi$ and $\E$, which leads to the Ernst equations
\begin{eqnarray}
   \nonumber
   &&f\cdot\left[(1-x^2)\E_{,xx}-2x\E_{,x}-(1-y^2)\E_{,yy}+2y\E_{,y}\right]\\
   &&   \qquad = (1-x^2)(\E_{,x}-2\bar\Phi\Phi_{,x})\E_{,x}
       -(1-y^2)(\E_{,y}-2\bar\Phi\Phi_{,y})\E_{,y},\label{eq:Ernst1}\\
   \nonumber
   &&f\cdot\left[(1-x^2)\Phi_{,xx}-2x\Phi_{,x}
          -(1-y^2)\Phi_{,yy}+2y\Phi_{,y}\right]\\
   &&   \qquad = (1-x^2)(\E_{,x}-2\bar\Phi\Phi_{,x})\Phi_{,x}
       -(1-y^2)(\E_{,y}-2\bar\Phi\Phi_{,y})\Phi_{,y}.\label{eq:Ernst2}
\end{eqnarray}
For more details we refer to \cite{Hennig2016a}.


A remarkable property of the \emph{nonlinear} Ernst equations \eqref{eq:Ernst1}, \eqref{eq:Ernst2} is that they are equivalent to an associated \emph{linear} problem (LP) via its integrability condition. 

Extending earlier work for the Einstein vacuum equations, an LP for the Einstein-Maxwell equations was first constructed by Belinski \cite{Belinski1979}. 
A different formulation of the LP in terms of a $3\times 3$ matrix $\mathbf\Omega$ is due to Neugebauer and Kramer \cite{NeugebauerKramer}. Here, similarly to a transformation considered by Meinel \cite{Meinel2012}, we use a minor reformulation of Neugebauer and Kramer's LP in terms of a  matrix $\Y$, which differs from $\mathbf\Omega$ only by multiplication by a diagonal matrix,
\begin{equation}
 \Y=\mathrm{diag}\left(1,1,-\frac{1}{\sqrt{f}}\right){\bf\Omega}.
\end{equation}
The advantage of this transformation is a simplification of some of the following calculations.

In order to formulate the LP, we first introduce coordinates $\mu$, $\nu$, which are defined in terms of $t$ and $\theta$ by
\begin{equation}
 \mu=\cos(t-\theta),\quad \nu=\cos(t+\theta).
\end{equation}
We also define the function
\begin{equation}\label{eq:lambda}
 \lambda(K, \mu,\nu)=\sqrt{\frac{K-\nu}{K-\mu}}, 
\end{equation}
which depends on the coordinates and on the \emph{spectral parameter} $K\in\CC$.

In terms of the above ingredients, the LP adapted to our situation reads
\begin{eqnarray}
 \Y_{,\mu} &=& \left[
  \left(
  \begin{array}{ccc}
   B_\mu & 0 & C_\mu\\ 0 & A_\mu & 0\\ -D_\mu & 0 & 0
  \end{array}\right)
   +\lambda
   \left(
   \begin{array}{ccc}
    0 & B_\mu & 0\\ A_\mu & 0 & -C_\mu\\ 0 & -D_\mu & 0
   \end{array}\right)
  \right]\Y,\\
  \Y_{,\nu} &=& \left[
  \left(
  \begin{array}{ccc}
   B_\nu & 0 & C_\nu\\ 0 & A_\nu & 0\\ -D_\nu & 0 & 0
  \end{array}\right)
   +\frac{1}{\lambda}
   \left(
   \begin{array}{ccc}
    0 & B_\nu & 0\\ A_\nu & 0 & -C_\nu\\ 0 & -D_\nu & 0
   \end{array}\right)
  \right]\Y,  
\end{eqnarray}
with
\begin{equation}
 A_\mu=\frac{\E_{,\mu}-2\bar\Phi\Phi_{,\mu}}{2f},\quad
 B_\mu=\frac{{\bar\E}_{,\mu}-2\Phi{\bar\Phi}_{,\mu}}{2f},\quad
 C_\mu=\Phi_{,\mu},\quad
 D_\mu=\frac{1}{f}{\bar\Phi}_{,\mu}
\end{equation}
and analogous definitions of $A_\nu$, $B_\nu$, $C_\nu$, $D_\nu$ in terms of $\nu$-derivatives. The integrability condition $\Y_{,\mu\nu}=\Y_{,\nu\mu}$ leads exactly to the Ernst equations \eqref{eq:Ernst1}, \eqref{eq:Ernst2}.

Similarly to the discussions of vacuum SGGTN solutions in \cite{BeyerHennig2012}, it is useful to consider the LP not only in the present coordinate system, but also in a frame with
\begin{equation}
 \tilde t=t,\quad \tilde\theta=\theta,\quad\tilde\rho_1=\rho_1+q\rho_2,\quad
 \tilde\rho_2=\rho_2,
\end{equation}
where $q=\mathrm{constant}$. This transformation, which corresponds to a rotation of the Killing basis, leaves the form of the metric invariant (and consequently also the form of the Ernst equations and the LP), but the potential $Q$ changes to
\begin{equation}
 \tilde Q=Q-q.
\end{equation}
Since $Q$ takes on the boundary values $\pm1$ on the axes, we will later consider the particular transformations with $q=\pm1$. Then we achieve that $\tilde Q=0$ on one of the axes, which leads to simplifications of the calculations. Note that there is a simple relationship between  $\Y$ and the matrix $\tilde\Y$ corresponding to the LP in the new coordinates \cite{HennigAnsorg2009,AnsorgHennig2009}. In our formulation, we have
\begin{equation}\label{eq:transf}
 \tilde\Y=\left[
 \left(\begin{array}{ccc}
        c_- & 0 & 0\\ 0 & c_+ & 0\\ 0 & 0 & 1
       \end{array}\right)
 +\ii(K-\mu)\frac{q}{f}
 \left(\begin{array}{ccc}
        1 & \lambda & 0\\ -\lambda & -1 & 0\\ 0 & 0 & 0
       \end{array}\right)
 \right]\Y,
\end{equation}
where
\begin{equation}
 c_{\pm}=1-q\left(a\pm\frac{\ii}{f}\sin t\sin\theta\right).
\end{equation}

\section{Integration of the linear problem on the boundaries\label{sec:integration}}

In the following we solve the linear problem along the axes and the past and future Cauchy horizons and continuously connect these solutions. Ultimately, these results will allow us to derive the Ernst potentials, as well as the metric and electromagnetic potentials, on the axes and the future horizon in terms of the data on the past horizon.

In our constructions, it is important that the function $\lambda$ [cf.~\eqref{eq:lambda}], considered as a function of the spectral parameter $K$ at a fixed coordinate position, has two possible values  for each $K\in\CC$ (due to the two possible signs of the square root), i.e.\ it is defined on a 2-sheeted Riemannian $K$-surface. Only at the branch points $K=\nu$ and $K=\mu$ (where $\lambda=0$ or $\lambda=\infty$, respectively) $\lambda$ is unique. This property must also show up in the behaviour of the matrix $\Y$, which can take on different values on the two sheets, but is unique at the branch points.

Note that, if $\Y$ is a \emph{particular} solution to the LP, then the most general solution is obtained by multiplication on the right by a matrix that depends on $K$ only, i.e.\ it is of the form $\Y{\bf C}(K)$. Moreover, if $\Y$ solves the LP on one sheet, then $\J\Y$ with 
\begin{equation}\label{eq:defJ}
 \J=\mathrm{diag}(1,-1,1)
\end{equation}
is a particular solution on the other sheet. The most general solution on the other sheet is then obtained through multiplication by a matrix $\B(K)$. Hence there must be a particular matrix $\B(K)$ such that the `correct' solution on the other sheet (the one that smoothly connects with the given $\Y$ through the branch cut) is of the form
\begin{equation}\label{eq:sheets}
 \Y(x,y,-\lambda)=\J\Y(x,y,\lambda)\B(K).
\end{equation}

The function $\lambda$ simplifies at the boundaries $t=0,\pi$ and $\theta=0,\pi$ of the Gowdy square, where we have $\mu=\nu$, in which case $\lambda=\pm1$. 

\subsection{Past horizon $\Hp$}

For $t=0$ (i.e.\ $y=1$) we obtain $\mu=\nu=\cos\theta=x$. We first solve the LP for $\lambda=1$. The solution for $\lambda=-1$ can then be obtained from \eqref{eq:sheets}. In this situation, the LP reduces to the ODE
\begin{equation}
 \Y_{,x}=\left(\begin{array}{ccc}
                B_\mu & B_\mu & C_\mu\\
                A_\mu & A_\mu & -C_\mu\\
                -D_\mu& -D_\mu& 0
               \end{array}\right)\Y.
\end{equation}
From the components of this equation, using the definitions of $A_\mu,\dots,D_\mu$, we easily obtain the general solution in terms of the Ernst potentials,
\begin{equation}\label{eq:defE}
 \lambda=1:\qquad
 \Y=\EE\C\p,\quad
 \EE:=\left(\begin{array}{ccc}
            \bar\E-2|\Phi|^2 & 1 & \Phi\\
            \E & -1 & -\Phi\\
            -2\bar\Phi & 0 & 1
           \end{array}\right),
\end{equation}
where the $3\times3$ matrix $\C\p=\C\p(K)$ is an integration `constant'.

Using \eqref{eq:transf} with $a=0$ [cf.~\eqref{eq:a}], we find the solution in the `rotated coordinates',
\begin{equation}
 \lambda=1: \qquad
 \tilde\Y=\left[\EE+2\ii q(K-\mu)\left(\begin{array}{ccc}
                                       1 & 0 & 0\\ -1 & 0 & 0\\ 0 & 0 & 0
                                      \end{array}\right)
   \right]\C\p.
\end{equation}

From \eqref{eq:sheets} we finally obtain on the other Riemannian sheet,
\begin{equation}\label{eq:Hpsheet2}
 \lambda=-1:\qquad
 \Y=\J\EE\C\p\B
\end{equation}
with a $3\times3$ matrix $\B=\B(K)$. Note that the rotated matrix $\tilde\Y$ is not required in the sheet $\lambda=-1$, for this does not add any new information.

\subsection{Axis $\Al$}

Following the same procedure, we solve the LP for $\theta=0$ ($x=1$), starting on the sheet $\lambda=1$. Again we obtain an ODE. From the solution we also construct $\tilde\Y$ (where we now choose $q=1$ due to $Q=1$ on $\Al$, and where we use $a=1/Q=1$) and $\Y$ on the other sheet. The results are
\begin{eqnarray}
 \fl\lambda=1:\quad && \Y=\EE\C_1,\\
 \fl                 && \tilde\Y=\left[
                \left(\begin{array}{ccc}
                        0 & 0 & 0\\ 0 & 0 & 0\\ 0 & 0 & 1
                       \end{array}\right)
               \EE+2\ii (K-\mu)\left(\begin{array}{ccc}
                                       1 & 0 & 0\\ -1 & 0 & 0\\ 0 & 0 & 0
                                      \end{array}\right)\right]\C_1,\\
 \fl \lambda=-1:\qquad && \Y=\J\EE\C_1\B,
\end{eqnarray}
with another integration `constant' $\C_1=\C_1(K)$, which we will later express in terms of $\C\p$ using continuity at $t=\theta=0$.

\subsection{Axis $\Ar$}

Here we have $Q=-1$ and $a=1/Q=-1$, and hence we choose $q=-1$. Then we obtain
\begin{eqnarray}
 \fl\lambda=1:\quad && \Y=\EE\C_2,\\
 \fl                 && \tilde\Y=\left[
                \left(\begin{array}{ccc}
                        0 & 0 & 0\\ 0 & 0 & 0\\ 0 & 0 & 1
                       \end{array}\right)
               \EE-2\ii (K-\mu)\left(\begin{array}{ccc}
                                       1 & 0 & 0\\ -1 & 0 & 0\\ 0 & 0 & 0
                                      \end{array}\right)\right]\C_2,\\
 \fl \lambda=-1:\qquad && \Y=\J\EE\C_2\B,
\end{eqnarray}
with $\C_2=\C_2(K)$.

\subsection{Future horizon $\Hf$}

Finally, we tentatively solve the LP at $t=\pi$. If we obtain a regular solution that can be continuously attached to the solutions on the axes, then the resulting spacetime has a regular future Cauchy horizon. Moreover, we will then be able to construct the potentials there in terms of the initial data. We will shortly see that regular solutions are indeed obtained in general, apart from special singular cases.

At this boundary, we have $a=a\f=\mathrm{constant}$ (the value of the constant will be determined later), and we find
\begin{eqnarray}
 \fl\lambda=1:\quad && \Y=\EE\C\f,\\
 \fl                 && \tilde\Y=\left[
                \left(\begin{array}{ccc}
                        1-qa\f & 0 & 0\\ 0 & 1-qa\f & 0\\ 0 & 0 & 1
                       \end{array}\right)
               \EE+2\ii (K-\mu)\left(\begin{array}{ccc}
                                       1 & 0 & 0\\ -1 & 0 & 0\\ 0 & 0 & 0
                                      \end{array}\right)\right]\C\f,\\
 \fl \lambda=-1:\qquad && \Y=\J\EE\C\f\B,
\end{eqnarray}
with $\C\f=\C\f(K)$.

\subsection{Continuity conditions}

We will see that the requirement of continuous transitions from $\Hp$ to $\Al$ and $\Ar$, and from these axes to $\Hf$, allows us to express the integration `constants' $\C_1$, $\C_2$ and $\C\f$ in terms of $\C\p$.

Firstly, we consider continuity of $\Y$ and $\tilde\Y$ (with $q=1$) at the point 
\begin{equation}
 A:\quad t=0,\ \theta=0\ (x=y=1),
\end{equation}
where $\mu=\nu=1$ and $f=0$. Then $\E\equiv \E_A=|\Phi_A|^2+\ii b_A$. Continuity of $\Y$ on the sheet $\lambda=1$ requires
\begin{equation}
 \EE\C\p=\EE\C_1\quad\textrm{at } A.
\end{equation}
The second rows of this matrix condition are the negatives of the first rows. Hence we obtain independent conditions only from the first and third rows. If these conditions are satisfied, then $\Y$ on the other sheet (with $\lambda=-1$) is automatically continuous at $A$.

Secondly, continuity of $\tilde\Y$ (with $q=1$ and $\lambda=1$) at $A$ leads to
\begin{eqnarray}
 \left[\EE+2\ii(K-1)\left(\begin{array}{ccc}
                           1 & 0 & 0\\
                           -1& 0 & 0\\
                           0 & 0 & 0
                          \end{array}\right)\right]\C\p\nonumber\\
 \quad=\left[\left(\begin{array}{ccc}
               0 & 0 & 0\\
               0 & 0 & 0\\
               0 & 0 & 1
              \end{array}\right)
 \EE+2\ii(K-1)\left(\begin{array}{ccc}
                           1 & 0 & 0\\
                           -1& 0 & 0\\
                           0 & 0 & 0
                          \end{array}\right)\right]\C_1.
\end{eqnarray}
Here, the second row conditions are again equivalent to the first rows, and the third row conditions are identical with the third row conditions from continuity of $\Y$. Hence we obtain one new row of conditions.

Using the definition \eqref{eq:defE} of $\EE$ and the form of the Ernst potentials at point $A$, the above conditions can be combined as follows,
\begin{equation}
 \fl
 \left(\begin{array}{ccc}
        \E_A & -1 & -\Phi_A\\
        \E_A-2\ii(K-1) & -1 & -\Phi_A\\
        -2\bar\Phi_A & 0 & 1
       \end{array}\right)\C\p
 =\left(\begin{array}{ccc}
        \E_A & -1 & -\Phi_A\\
        -2\ii(K-1) & 0 & 0\\
        -2\bar\Phi_A & 0 & 1
       \end{array}\right)\C_1.
\end{equation}
Since the first matrix on the right-hand side is invertible, we can easily solve this equation for $\C_1$. The result is
\begin{equation}\label{eq:C1}
 \C_1=\left(\1+\frac{1}{2\ii(K-1)}\M_A\right)\C\p,
\end{equation}
where $\1$ is the $3\times3$ identity matrix and
\begin{equation}\label{eq:defMA}
 \M_A=m_A n_A^T,\quad
 m_A:=\left(\begin{array}{c}
             1\\ -\bar\E_A\\ 2\bar\Phi_A
            \end{array}\right),\quad
 n_A:=\left(\begin{array}{c}
             -\E_A\\ 1\\ \Phi_A
            \end{array}\right).           
\end{equation}
Note that the matrix $\M_A$ has the useful property $\M_A^2=\0$. This follows from $m_A\cdot n_A\equiv m_A^Tn_A=-2f_A=0$.

Next, we consider continuity at 
\begin{equation}
 B:\quad t=0,\ \theta=\pi\ (x=-1,\ y=1).
\end{equation}
With a similar calculation as above, we now obtain a formula for $\C_2$ in terms of $\C\p$,
\begin{equation}\label{eq:C2}
 \C_2=\left(\1-\frac{1}{2\ii(K+1)}\M_B\right)\C\p,
\end{equation}
where the definition of $\M_B$ is obtained from \eqref{eq:defMA} by replacing the subscript $A$ with $B$.

Finally, we study the transition from the axes to the future horizon and look at continuity at the points
\begin{equation}
 C:\quad t=\pi,\ \theta=0\ (x=1,\ y=-1)
\end{equation}
and
\begin{equation}
 D:\quad t=\pi,\ \theta=\pi\ (x=y=-1).
\end{equation}
With our formulae for $\C_1$ and $\C_2$, continuity at $C$ leads to
\begin{equation}\label{eq:Cf}
 \C\f=\left(\1-\frac{1-a\f}{2\ii(K+1)}\M_C\right)
      \left(\1+\frac{1}{2\ii(K-1)}\M_A\right)\C\p,
\end{equation}
while continuity at $D$ gives
\begin{equation}\label{eq:Cf2}
 \C\f=\left(\1+\frac{1+a\f}{2\ii(K-1)}\M_D\right)
      \left(\1-\frac{1}{2\ii(K+1)}\M_B\right)\C\p,
\end{equation}
with the obvious definitions of $\M_C$ and $\M_D$. Since there is just one matrix $\C\f$, the right-hand sides of the two previous formulae must be the same. Equating these and multiplying by $(K+1)(K-1)$, we obtain the condition that two matrix polynomials in the spectral parameter $K$ must agree. Comparing coefficients of different powers of $K$, we obtain a system of equations that can be solved for the vectors $m_C$, $n_C$, $m_D$, $n_D$ and the constant $a\f$ (the value of the auxiliary function $a$ at $\Hf$). The solutions are
\begin{equation}
 m_C=\frac{(m_B\cdot n_A)m_A-4\ii m_B}{m_B\cdot n_A-4\ii},\quad
 n_C=\frac{(m_A\cdot n_B)n_A+4\ii n_B}{m_A\cdot n_B+4\ii},
\end{equation}
\begin{equation}
 m_D=\frac{(m_A\cdot n_B)m_B-4\ii m_A}{m_A\cdot n_B-4\ii},\quad
 n_D=\frac{(m_B\cdot n_A)n_B+4\ii n_A}{m_B\cdot n_A+4\ii}
\end{equation}
and 
\begin{equation}
 a\f = -\frac{8\,\Im(m_A\cdot n_B)}{|m_A\cdot n_B|^2+16}.
\end{equation}
Since the components of the vectors $m$ and $n$ are defined in terms of the Ernst potentials, we can reformulate these equations and obtain expressions for $a\f$ and for the values of the Ernst potentials at $C$ and $D$ in terms of the data at the past horizon,
\begin{equation}\label{eq:Cvals}
 \fl
 \E_C=\frac{4\ii\E_B-(\bar\E_A+\E_B-2\bar\Phi_A\Phi_B)\E_A}
           {4\ii-(\bar\E_A+\E_B-2\bar\Phi_A\Phi_B)},\quad
 \Phi_C=\frac{4\ii\Phi_B-(\bar\E_A+\E_B-2\bar\Phi_A\Phi_B)\Phi_A}
             {4\ii-(\bar\E_A+\E_B-2\bar\Phi_A\Phi_B)},
\end{equation}
\begin{equation}\label{eq:Dvals}
 \fl
 \E_D=\frac{4\ii\E_A-(\E_A+\bar\E_B-2\Phi_A\bar\Phi_B)\E_B}
           {4\ii- (\E_A+\bar\E_B-2\Phi_A\bar\Phi_B)},\quad
 \Phi_D=\frac{4\ii\Phi_A-(\E_A+\bar\E_B-2\Phi_A\bar\Phi_B)\Phi_B}
           {4\ii- (\E_A+\bar\E_B-2\Phi_A\bar\Phi_B)},
\end{equation}
and
\begin{equation}\label{eq:af}
 a\f=-8\frac{\Im(\E_A+\bar\E_B-2\Phi_A\bar\Phi_B)}{|\E_A+\bar\E_B-2\Phi_A\bar\Phi_B|^2+16}.
\end{equation}
Note that the values of the Ernst potentials are well-defined, unless $\bar\E_A+\E_B-2\bar\Phi_A\Phi_B=\pm4\ii$, in which case $\E_C$ or $\E_D$ diverge (while $\Phi_C$ and $\Phi_D$ remain bounded as the zero of the denominator is compensated by a zero of the numerator). These `singular cases' will be discussed in more detail below.

\subsection{Ernst potentials at the boundaries}

With our solutions of the LP at the boundaries, we are now in a position to find expressions for the Ernst potentials at the axes and the future horizon in terms of the Ernst potentials at the past horizon, i.e.\ the initial data.

We use that the matrix $\Y$ is defined on the 2-sheeted Riemannian $K$-surface. As remarked above, $\Y$ can generally take on different values on the two sheets, but must be unique at the two branch points $K=\mu$ and $K=\nu$. As we approach the boundaries of the Gowdy square, the two branch points come closer and closer. At the boundaries, where $\mu=\nu$ holds, we have confluent branch points at $K=\mu=\nu$. The requirement that the values of $\Y$ on the two sheets with $\lambda=\pm1$ agree at $K=\mu=\nu$ allows us to construct the desired relations for the Ernst potentials.

At the past horizon $\Hp$, the uniqueness condition for $\Y$ reads [cf.~\eqref{eq:defE}, \eqref{eq:Hpsheet2}]
\begin{equation}\label{eq:Hpcond}
 \EE\p\C\p = \J\EE\p\C\p\B\quad\textrm{at}\quad K=\mu=\nu=x.
\end{equation}
This condition fixes some of the degrees of freedom of the matrix $\C\p$ at $K=x$ in terms of the initial data. The remaining degrees of freedom do not have any physical relevance and could be fixed by imposing certain gauge conditions. Here, however, we will keep the general form of $\C\p$, since the formulae for the Ernst potentials will be independent of any gauge choice.

Similarly, the uniqueness condition at the axis $\Al$ is
\begin{equation}\label{eq:A1cond}
 \EE_1\C_1 = \J\EE_1\C_1\B\quad\textrm{at}\quad K=\mu=\nu=y,
\end{equation}
where $\C_1$ is given in terms of $\C\p$ by our earlier result \eqref{eq:C1}. Hence \eqref{eq:A1cond} is a condition for $\EE_1$, the components of which will provide us with the Ernst potentials at $\Al$. In order to solve for the Ernst potentials, we proceed as follows.

\begin{figure}
 \centering
 \includegraphics[width=14cm]{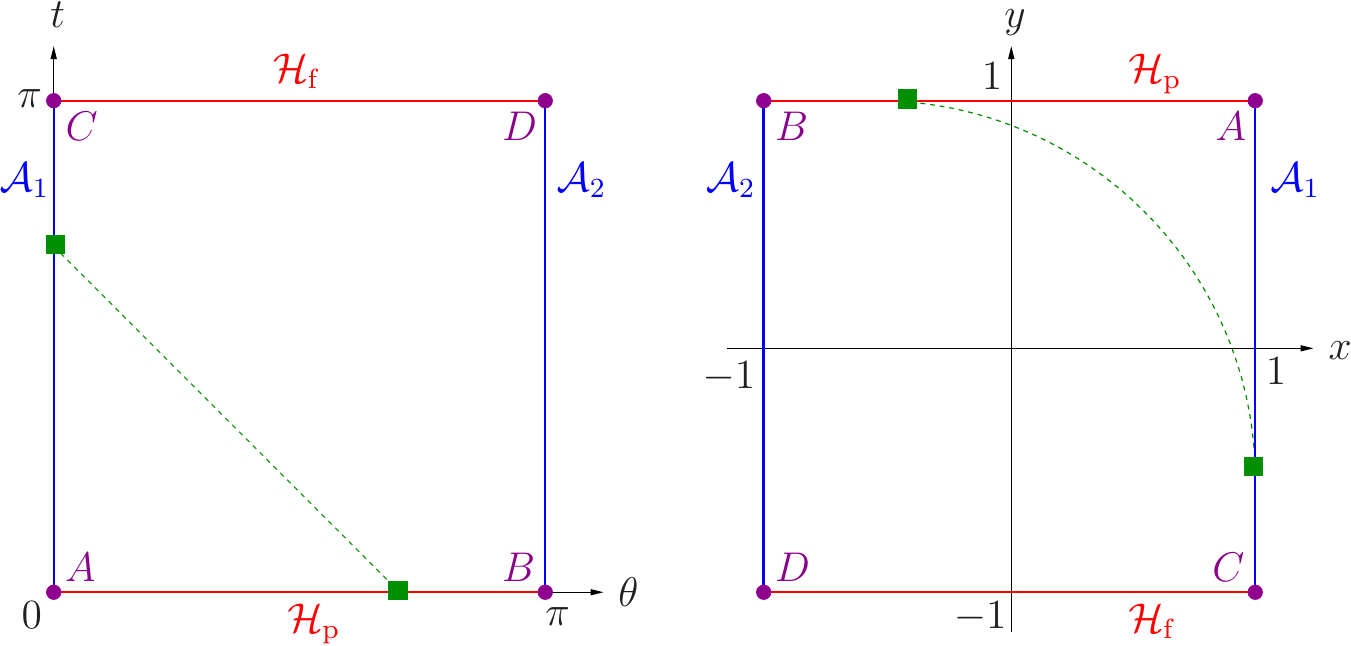}
 \caption{The Gowdy square is shown with respect to the coordinates $(\theta,t)$ and $(x,y)$. The boundaries are the axes $\Al$, $\Ar$, and the past and future Cauchy horizons $\Hp$, $\Hf$, respectively. The values of the Ernst potentials at the marked point on $\Al$, which are given by Eqns.~\eqref{eq:E1} and \eqref{eq:Phi1}, depend on the values of the initial Ernst potentials at point $A$ and at that point on $\Hp$ that is connected to the point on $\Al$ via a null geodesic. (Null geodesics are simple straight lines in the $\theta$-$t$-plane, and elliptical arcs in the $x$-$y$-plane.)
 \label{fig:Gowdy-square}}
\end{figure}

We define a matrix 
\begin{equation}
 \H = \left(\begin{array}{ccc} 
             h_1 & h_2 & h_3\\ h_4 & h_5 & h_6\\ h_7 & h_8 & h_9              
            \end{array}\right)
\end{equation}
by writing $\EE_1$ as
\begin{equation}\label{eq:E1eq}
 \EE_1=:\H\EE\p\left(\1-\frac{1}{2\ii(\mu-1)}\M_A\right).
\end{equation}
Plugging this expression into \eqref{eq:A1cond} and using that $\M_A^2=\0$, we obtain
\begin{equation}
 \H\EE\p\C\p = \J\H\EE\p\C\p\B.
\end{equation}
Now we replace $\EE\p\C\p$ on the left-hand side using \eqref{eq:Hpcond} and arrive at the condition
\begin{equation}
 \H = \J\H\J.
\end{equation}
With the definition of $\J$ in \eqref{eq:defJ}, we see that this implies
\begin{equation}
 h_2=h_4=h_6=h_8=0.
\end{equation}
Hence Eq.~\eqref{eq:E1eq} becomes
\begin{equation}\label{eq:E1eq2}
 \EE_1=\left(\begin{array}{ccc}
              h_1 & 0 & h_3\\ 0 & h_5 & 0\\ h_7 & 0 & h_9
             \end{array}\right)
       \EE\p\left(\1-\frac{1}{2\ii(\mu-1)}m_An_A^T\right).
\end{equation}
The nine components of this condition provide us with nine equations for the nine unknowns $h_1$, $h_3$, $h_5$, $h_7$, $h_9$, $\E_1$, $\bar\E_1$, $\Phi_1$, $\bar\Phi_1$ appearing in this equation. Since we are only interested in the Ernst potentials, it is sufficient to consider the second row of \eqref{eq:E1eq2}. Of the functions $h_1,\dots,h_9$, only $h_5$ appears in this row, and the ratios of two of the equations in this row are independent of $h_5$ and can be solved for $\E_1$ and $\Phi_1$. The results are
\begin{eqnarray}
 \E_1(y) &=& \left.\frac{(\E\p+\bar\E_A-2\Phi\p\bar\Phi_A)\E_A-2\ii(1-y)\E\p}
                  {(\E\p+\bar\E_A-2\Phi\p\bar\Phi_A)-2\ii(1-y)}\right|_{x=y},
                  \label{eq:E1}\\
 \Phi_1(y) &=& \left.\frac{(\E\p+\bar\E_A-2\Phi\p\bar\Phi_A)\Phi_A-2\ii(1-y)\Phi\p}
                  {(\E\p+\bar\E_A-2\Phi\p\bar\Phi_A)-2\ii(1-y)}\right|_{x=y}.
                  \label{eq:Phi1}
\end{eqnarray}
Note that the initial data $\E\p$ and $\Phi\p$ appear on the right-hand sides, which are functions of $x$. This argument, however, needs to be replaced by $y$ to find $\E_1$ and $\Phi_1$ as functions of $y$. This comes from the fact that $\mu=x$ on $\Hp$ and $\mu=y$ on $\Al$. Hence the values of the initial data at two points on $\Hp$ (the point $A$ and the point $x=y$) are required to fix the Ernst potentials at the point $y$ on the axis $\Al$.
The relevant points are shown in Fig.~\ref{fig:Gowdy-square}.
On the other hand, since the relationship between the Ernst and metric potentials is nonlocal and involves first-order ODEs, the metric at the point $y$ on the axis $\Al$ also requires the initial metric functions \emph{everywhere between} those two points on $\Hp$. 

Next we consider the uniqueness condition at the other axis $\Ar$, which reads
\begin{equation} \label{eq:A2cond}
 \EE_2\C_2 = \J\EE_2\C_2\B\quad\textrm{at}\quad K=\mu=\nu=-y,
\end{equation}
with $\C_2$ from \eqref{eq:C2}. Without repeating the previous calculations, we simply observe that this condition and our earlier condition \eqref{eq:A1cond} can be translated into each other with the replacements
\begin{equation}
 1\leftrightarrow 2,\quad
 A\leftrightarrow B,\quad
 \mu\leftrightarrow -\mu.
\end{equation}
Hence we can directly obtain the axis potentials $\E_2$, $\Phi_2$ from the above formulae for $\E_1$, $\Phi_1$ with the same replacements. We find
\begin{eqnarray}
 \E_2(y) &=& \left.\frac{(\E\p+\bar\E_B-2\Phi\p\bar\Phi_B)\E_B-2\ii(1-y)\E\p}
                  {(\E\p+\bar\E_B-2\Phi\p\bar\Phi_B)-2\ii(1-y)}\right|_{x=-y},
                  \label{eq:E2}\\
 \Phi_2(y) &=& \left.\frac{(\E\p+\bar\E_B-2\Phi\p\bar\Phi_B)\Phi_B-2\ii(1-y)\Phi\p}
                  {(\E\p+\bar\E_B-2\Phi\p\bar\Phi_B)-2\ii(1-y)}\right|_{x=-y}.
                  \label{eq:Phi2}
\end{eqnarray}

Finally, we calculate the Ernst potentials on the future horizon $\Hf$. The calculations follow the same steps as before, but they are slightly more complicated since the formula \eqref{eq:Cf} for $\C\f$ is more involved than the equations for $\C_1$ and $\C_2$. For a concise formulation of the results, we first define seven constants,
\begin{equation}
 \gamma_0 = |m_A\cdot m_B|^2+16\equiv |\E_A+\bar\E_B-2\Phi_A\bar\Phi_B|^2+16,
\end{equation}
\begin{eqnarray}
\fl
 \left(\begin{array}{c}\gamma_1\\ \gamma_2\\ \gamma_3\end{array}\right)
  &\!\!=& (m_B\cdot n_A)n_B+4\ii n_A
  \equiv -(\E_A+\bar \E_B-2\Phi_A\bar\Phi_B)
      \left(\begin{array}{c}-\E_B\\ 1\\ \Phi_B\end{array}\right)
      +4\ii\left(\begin{array}{c}-\E_A\\ 1\\ \Phi_A\end{array}\right)\!,\\
\fl
 \left(\begin{array}{c}\gamma_4\\ \gamma_5\\ \gamma_6\end{array}\right)
  &\!\!=& (m_A\cdot n_B)n_A+4\ii n_B
  \equiv -(\E_B+\bar \E_A-2\Phi_B\bar\Phi_A)
      \left(\begin{array}{c}-\E_A\\ 1\\ \Phi_A\end{array}\right)
      +4\ii\left(\begin{array}{c}-\E_B\\ 1\\ \Phi_B\end{array}\right)\!.
\end{eqnarray}
We also define the following two functions,
\begin{eqnarray}
 F_A(x) &=& \E\p(x)-\E_A-2\bar\Phi_A[\Phi\p(x)-\Phi_A],\\
 F_B(x) &=& \E\p(x)-\E_B-2\bar\Phi_B[\Phi\p(x)-\Phi_B].
\end{eqnarray}
In terms of these abbreviations, the Ernst potentials at $\Hf$ can be written as
\begin{eqnarray}
 \fl\label{eq:EPf1}
 \E\f(x) &=& \frac{\gamma_0(1-x^2)\E\p(-x)-2\gamma_1(1-x)F_A(-x)-2\gamma_4(1+x)F_B(-x)}
               {\gamma_0(1-x^2)+2\gamma_2(1-x)F_A(-x)+2\gamma_5(1+x)F_B(-x)},\\
 \fl\label{eq:EPf2}
 \Phi\f(x) &=& \frac{\gamma_0(1-x^2)\Phi\p(-x)+2\gamma_3(1-x)F_A(-x)+2\gamma_6(1+x)F_B(-x)}
               {\gamma_0(1-x^2)+2\gamma_2(1-x)F_A(-x)+2\gamma_5(1+x)F_B(-x)}.
\end{eqnarray}

\subsection{Regularity of the axes potentials}

We have seen that, starting from regular data at the past horizon $\Hp$, we can integrate the linear problem along the boundaries $\Al$, $\Ar$, $\Hp$, $\Hf$ and obtain a regular solution --- provided the constants $\E_C$, $\E_D$, $\Phi_C$, $\Phi_D$ and $a\f$ that appear in the solution are finite. As mentioned above, this is generally true. The only exceptions are initial Ernst potentials satisfying
\begin{equation}\label{eq:singcon1}
 \bar\E_A+\E_B-2\bar\Phi_A\Phi_B=\pm4\ii.
\end{equation}
In those cases, the denominators in \eqref{eq:Cvals} or \eqref{eq:Dvals} vanish, while the numerators in the formulae for $\E_C$ or $\E_D$ are nonzero (even though the numerators of $\Phi_C$ or $\Phi_D$ vanish as well).
Note that the condition \eqref{eq:singcon1} can be written as
\begin{eqnarray}
 -\ii b_A+\Phi_A\bar\Phi_A+\ii b_B+\Phi_B\bar\Phi_B -2\bar\Phi_A\Phi_B\nonumber\\
 \equiv|\Phi_A-\Phi_B|^2+\ii[b_B-b_A+2\Im((\Phi_A-\Phi_B)\bar\Phi_A)]=\pm4\ii.
\end{eqnarray}
Hence the real and imaginary parts on both sides are equal if and only if the initial Ernst potentials satisfy
\begin{equation}\label{eq:singcon2}
 b_B-b_A=\pm 4\quad\textrm{and}\quad \Phi_A=\Phi_B.
\end{equation}
In the following we will see that the case of a plus (minus) sign corresponds to a solution with a singularity at point $C$ (point $D$), and that the points $C$ and $D$ are the only potentially singular points in the Gowdy square.

Firstly, we can verify that the axes Ernst potentials are always regular for $t<\pi$. This is already guaranteed by the general theory, which ensures regularity in the region $t<\pi$, but can be seen from the explicit expressions as well. Since the axes potentials are rational functions of the initial potentials, which are assumed to be regular, a singularity could only appear if the denominator vanishes. It quickly follows that the real part of the denominator could only vanish if $f\p(x)$ vanishes at $x=y$ (on $\Al$) or at $x=-y$ (on $\Ar$) for some $y\in(-1,1]$. Since $f\p(x)=0$ does only hold for $x=\pm1$, and $f\p(x)>0$ otherwise [cf.\ \eqref{eq:f}], the denominator could only vanish at $y=\pm1$.
Since we currently assume $t<\pi$, i.e.\ $y>-1$, only the case $y=1$ remains, and it turns out that the denominator does vanish there. However, the numerator vanishes there as well, and the quotient remains regular. (Indeed, we start from regular potentials at the corresponding points $A$ and $B$.) 

If we now include $t=\pi$ ($y=-1$), then the only other possibility for a singularity is that $y=-1$, corresponding to the points $C$ or $D$. There, as we already know, we can indeed have singularities. The expression for the denominator implies, however, that this can only happen in the above mentioned singular cases, i.e.\ for initial potentials satisfying \eqref{eq:singcon2}. Hence the axes potentials are generally regular, with the only exceptions being the singular cases, in which we have singularities at the points $C$ or $D$.

For regularity of the metric potentials, which can be constructed from the Ernst potentials, it is also necessary that the function $f$ does not vanish for $y\in(-1,1)$, cf.~\eqref{eq:f} with $x=\pm1$. Since $\E_1=f_1+|\Phi_1|^2+\ii b_1$, we have $f_1=\frac12(\E_1+\bar\E_1)-\Phi_1\bar\Phi_1$. With the formulae for $\E_1$, $\Phi_1$ and the complex conjugate expressions, we obtain
\begin{equation}
 f_1(y) = \frac{4(1-y)^2f\p(y)}{|\E\p(y)+\bar\E_A-2\bar\Phi_A\Phi\p(y)-2\ii(1-y)|^2}.
\end{equation}
[Note that the denominator is the squared modulus of the denominator in the formulae for $\E_1$ and $\Phi_1$, which, as we have already seen, cannot vanish for $y\in(-1,1)$.] Obviously, positivity of $f\p(x)$ for $x\in(-1,1)$ then implies positivity of $f_1(y)$ for $y\in(-1,1)$. 

A similar calculation at the other axis $\Ar$ shows that $f_2(y)$ is positive as well for $y\in(-1,1)$.

Finally, we have a closer look at the singular cases. Since the Ernst potentials are defined in terms of the Killing vectors, diverging potentials indicate a physical singularity rather than a coordinate effect. We can, however, also verify that other scalar quantities diverge in the singular cases. As an example, we consider the Kretschmann scalar $k=R_{ijkl}R^{ijkl}$ at $\Al$. 

We assume that initial data satisfying $b_B-b_A=4$ and $\Phi_A=\Phi_B$ are chosen, which give rise to a singularity at the point $C$. In this case, the denominator in the formula for $\E_1$ approaches zero as $y\to-1$, and the functions $f$ and $b$ have the leading-order behaviour
\begin{equation}
 f_1=\frac{f_0}{y+1}+\mathcal O(1),\quad
 b_1=\frac{b_0}{y+1}+\mathcal O(1),
\end{equation}
for some constants $f_0$ and $b_0$. The potential $\Phi_1$, on the other hand, remains regular as $y\to-1$, because the zero of the denominator is compensated by a zero of the numerator. With this behaviour of the potentials near $y=-1$, it follows from a (computer algebra) calculation that the Kretschmann scalar has the form
\begin{equation}
 k=2\frac{6f_0^2+b_0^2-8f_0^2b_0^2}{f_0^2R_0^2(y+1)^6}+\mathcal O[(y+1)^{-5}].
\end{equation}
Hence $k$ generally diverges like $(y+1)^{-6}$. However, for particular solutions for which $6f_0^2+b_0^2-8f_0^2b_0^2=0$ holds, the leading singular term vanishes and we obtain a weaker\footnote{In order to avoid possible misunderstandings, we point out that the singularity is not `weaker' in the sense that it becomes a different type of singularity, e.g., a quasi-regular singularity. It is still a \emph{scalar curvature singularity} for which curvature scalars (like the Kretschmann scalar $k$) diverge. It is `weaker'  in that $k$ diverges with a smaller inverse power of the coordinate distance to the singularity.} singularity. This is similar to the vacuum case, in which we observed that we can approach the singularity along a carefully chosen path for which the singularity is weaker, and there are even paths for which $k$ remains finite \cite{BeyerHennig2014}. Hence $k$ shows a directional behaviour. Here we have chosen a fixed path (the axis $\Al$), but we can fine-tune the initial data to achieve the same effect.

An analogous diverging behaviour of the Kretschmann scalar is observed on the other axis $\Ar$.

\subsection{Regularity at the Cauchy horizon}

Similarly to our investigation of regularity of the Ernst potentials at the axes in the previous subsection, the formulae for the potentials \eqref{eq:EPf1}, \eqref{eq:EPf2} at the future horizon $\Hf$ reveal that a singularity could only result from a zero of the denominators. However, instead of studying the rather lengthy expressions in these equations, it is much easier to use a reformulation. In \eqref{eq:EPf1} and \eqref{eq:EPf2}, the Ernst potentials at $\Hf$ are expressed in terms of the data at the past horizon $\Hp$. Alternatively, since we have already established regularity of the axes potentials, we can also construct and study formulae for $\E\f$ and $\Phi\f$ in terms of, say, $\E_1$ and $\Phi_1$ rather than $\E\p$ and $\Phi\p$. Following our above method for constructing $\E\f$ and $\Phi\f$, but expressing the matrix $\C\f$ in terms of $\C_1$ instead of $\C\p$, cf.\ \eqref{eq:C1} and \eqref{eq:Cf}, we obtain the potentials at the future horizon in terms of the potentials on $\Al$. The results is
\begin{equation}
 \E\f(x) = \left.\frac{(1-a\f)(\E_1+\bar\E_C-2\bar\Phi_C\Phi_1)\E_C-2\ii(1-x)\E_1}
             {(1-a\f)(\E_1+\bar\E_C-2\bar\Phi_C\Phi_1)-2\ii(1-x)}\right|_{y=-x},
\end{equation}
\begin{equation}
 \Phi\f(x) = \left.\frac{(1-a\f)(\E_1+\bar\E_C-2\bar\Phi_C\Phi_1)\Phi_C-2\ii(1-x)\Phi_1}
             {(1-a\f)(\E_1+\bar\E_C-2\bar\Phi_C\Phi_1)-2\ii(1-x)}\right|_{y=-x}.
\end{equation}
A necessary condition for a singularity is a zero of the denominator for some $x\in(-1,1)$. (Since we already discussed singularities at the points $C$ and $D$, corresponding to $x=\pm1$, we can restrict ourselves to the open interval.) Expressing the axes potentials in terms of their ingredients $f$, $b$, $A_4$ and $\beta$, the denominator can be written as
\begin{eqnarray}\label{eq:denom}
 (1-a\f)\left[f_1+\frac{1}{R_0}(A_{41}-A_{4C})^2+\frac{1}{R_0}(\beta_1-\beta_C)^2\right]\nonumber\\
 \quad +\ii\left[(1-a\f)\left(b_1-b_C-\frac{2}{R_0}(A_{4C}\beta_1-A_{41}\beta_C)\right)-2(1-x)\right].
\end{eqnarray}
The real part of \eqref{eq:denom} can only vanish if $a\f=1$ (since $f_1>0$ in the considered interval). With the formula \eqref{eq:af} for $a\f$, it follows that $a\f=1$ is only possible for $\Phi_A=\Phi_B$ and $b_B-b_A=4$, i.e.\ in one of the earlier discussed singular cases. The condition that the imaginary part of \eqref{eq:denom} vanishes as well implies that $x=1$ --- in contradiction to our assumption $x\in(-1,1)$. Hence there are no singularities in that interval. The only possible singularities occur at the points $C$ or $D$ in the earlier discussed singular cases. Thus the Ernst potentials in the interior horizon region are free of singularities.

Similarly to our investigation of regularity at the axes, regularity of the metric potentials constructed from the Ernst potentials requires, in addition, that $f>0$ on $\Hf$. We obtain from the Ernst potentials on $\Hf$ that
\begin{equation}
 f\f=\left.\frac{4(1-x)^2f_1}{|(1-a\f)(\E_1+\bar\E_C-2\bar\Phi_C\Phi_1)-2\ii(1-x)|^2}\right|_{y=-x}.
\end{equation}
With the previously shown positivity of $f_1$, this proves positivity of $f_f$.

Finally, we note that the solutions can be smoothly extended through the past and future Cauchy horizons. This follows from the same construction as discussed in \cite{BeyerHennig2014,Hennig2016a}: With respect to the coordinates $(x,y,\rho_1,\rho_2)$, the metric is obviously singular at $y=\pm1$, but these singularities are removable coordinate singularities. First we consider the general case with $a\f\neq 0$ [cf.\ \eqref{eq:af}]. Regular coordinates can be introduced with the transformation
\begin{equation}\label{eq:regcoords}
 \rho_1'=\rho_1-\kappa\ln(1-y)-\kappa_1\ln(1+y),
 \quad
 \rho_2'=\rho_2-\kappa_2\ln(1+y)
\end{equation}
with constants $\kappa$, $\kappa_1$, $\kappa_2$ such that\footnote{As a consequence of the Einstein-Maxwell equations, the expressions on the right-hand sides in \eqref{eq:kappa} are independent of $x$, i.e.\ $\kappa$, $\kappa_1$ and $\kappa_2$ are indeed constants as required.}
\begin{equation}\label{eq:kappa}
 \kappa^2=\lim_{y\to 1}\frac{\ee^{M-u}}{4R_0},\quad
 \kappa_2^2=\lim\limits_{y\to -1}\frac{(1-y^2)\ee^{M+u}}{4R_0(1-x^2)},\quad
 \kappa_1=-\kappa_2\lim_{y\to-1}Q.
\end{equation}
Note that we can independently select the signs of $\kappa$ and $\kappa_2$, which gives rise to different extensions\footnote{The four possible signs of the the two parameters correspond to four extensions. We expect that two of these are actually inequivalent, while the other two are isometric (and hence equivalent) to one of these. See the related discussion of extensions of an exact vacuum SGGTN solution in Sec.~4.6 of \cite{BeyerHennig2014}.}. Then the metric in the coordinates $(x,y,\rho_1',\rho_2')$ is regular at $y=\pm1$. Choosing smooth extensions of the metric potentials to $|y|>1$, we obtain extensions of the corresponding solution beyond the Cauchy horizons. 
Since $\rho_1'$ becomes a timelike coordinate in the regions with $|y|>1$, it follows from the periodicity of the $\rho$-coordinates that closed timelike curves exist there --- as expected behind Cauchy horizons.

In the special case $a\f=0$ (corresponding to initial data with $\E_A+\bar\E_B-2\Phi_A\bar\Phi_B=0$), the transformation \eqref{eq:regcoords} can still be used, but --- due to a different behaviour of the metric potentials at $\Hf$ --- we have to modify the formulae for the constants as follows~\cite{Hennig2016a},
\begin{equation}
 \kappa^2=\lim_{y\to 1}\frac{\ee^{M-u}}{4R_0},\quad
 \kappa_1^2=\lim_{y\to -1}\frac{\ee^{M-u}}{4R_0},\quad
 \kappa_2=0.
\end{equation}
In this case, we can choose arbitrary signs for $\kappa$ and $\kappa_1$.

\section{Discussion\label{sec:discussion}}

We have extended the class of smooth Gowdy-symmetric generalised Taub-NUT (SGGTN) solutions, which was first introduced in \cite{BeyerHennig2012}, from vacuum to electrovacuum. In this way, we have obtained a class of Gowdy-symmetric solutions  to the Einstein-Maxwell equations in electrovacuum with spatial $\Sth$-topology. The solutions have a regular past Cauchy horizon (generated by the Killing vector $\partial_{\rho_1}$) and, generally, develop a second, future Cauchy horizon (generated by a linear combination of $\partial_{\rho_1}$ and $\partial_{\rho_2}$). Moreover, they have four functional degrees of freedom (we can specify $u$, $Q$, $A_{3,y}$ and $A_4$ as smooth functions on the past horizon), which should be chosen subject to the necessary conditions for local existence \eqref{eq:conditions}. Compared to the vacuum case with two degrees of freedom ($u$ and $Q$), we have two additional degrees of freedom due to the electromagnetic field.

Properties of the new class of solutions have been derived with methods from soliton theory, namely from a discussion of the linear matrix problem that is equivalent to the Einstein-Maxwell equations (in the Ernst formulation) via its integrability condition. In particular, we have obtained explicit expressions for the potentials at the axes and the future horizon in terms of data at the past horizon [cf.\ Eqs.\ \eqref{eq:E1}, \eqref{eq:Phi1}; \eqref{eq:E2}, \eqref{eq:Phi2}; \eqref{eq:EPf1}, \eqref{eq:EPf2}].

As mentioned above, we find that the solutions are regular in the Gowdy square $0<t<\pi$ and, generally, also at $t=\pi$, where they have a future Cauchy horizon. Furthermore, the solutions can be extended as smooth solutions through the Cauchy horizons.
Only in special singular cases, obtained for initial values of the Ernst potential satisfying both conditions in \eqref{eq:singcon2}, a curvature singularity is formed at the points $C$ or $D$ ($t=\pi$ and $\theta=0$ or $\pi$). In that case, the Kretschmann scalar diverges at the singular point. 

Hence we observe that the Cauchy horizons of vacuum SGGTN solutions are stable with respect to particular electromagnetic perturbations, since the corresponding electromagnetic SGGTN solutions still have regular Cauchy horizons.

These results show that Gowdy-symmetric electromagnetic fields do not destroy the structure of the corresponding vacuum solutions, and the extended class of electromagnetic SGGTN solutions behaves very similarly to the vacuum SGGTN solutions.

\section*{Acknowledgments}
 We would like to thank Boris Daszuta for commenting on the manuscript. This research was supported by the University of Otago through a Division of Sciences Strategic Seeding Grant.
 


\end{document}